\documentclass{emulateapj}

\usepackage{amsmath}
\usepackage{array}
\usepackage{bm}
\usepackage{cases}
\usepackage{txfonts}
\usepackage{tipa}
\usepackage{amsfonts}
\usepackage{amssymb}
\usepackage{graphicx}
\usepackage{multirow}
\usepackage{float}
\usepackage{longtable}

\shortauthors{Du Cuihua et al.}

\begin{document}

	\title{New nearby hypervelocity stars and their spatial distribution from Gaia DR2 }
	
	\author{Cuihua Du\altaffilmark{1}, Hefan Li\altaffilmark{2}, Yepeng Yan\altaffilmark{2}, Heidi Jo Newberg\altaffilmark{3}, Jianrong Shi\altaffilmark{4,1}, Jun Ma\altaffilmark{4,1}, Yuqin Chen\altaffilmark{4,1}, Zhenyu  Wu\altaffilmark{4,1}}
	
	\affil{$^{1}$College of Astronomy and Space Sciences, University of Chinese Academy of Sciences, Beijing 100049, China; ducuihua@ucas.ac.cn\\
		$^{2}$School of Physical Sciences, University of Chinese Academy of Sciences, Beijing 100049,  China \\
		$^{3}$Department of Physics, Applied Physics and Astronomy, Rensselaer Polytechnic Institute, Troy, NY 12180, USA\\
		$^{4}$Key Laboratory of Optical Astronomy, National Astronomical Observatories, Chinese Academy of Sciences, Beijing 100012, China\\
	}

\begin{abstract}
\par Base on about 4,500 large tangential velocity ($V_\mathrm{tan}>0.75V_\mathrm{esc}$) with high-precision proper motions and $5\sigma$ parallaxes in 
Gaia DR2 5D information derived from parallax and proper motion, we identify more than 600 high velocity stars with $50\%$ unbound probability. Of these, 28 nearby (less than 6 kpc) late-type Hypervelocity stars (HVSs) with over $99\%$ possibility of unbound are discovered. In order to search for the unbound stars from the full Gaia DR2 6D phase space information derived from parallax, proper motion and radial velocity, we also identify 28 stars from the total velocity ($V_\mathrm{gc}>0.75V_\mathrm{esc}$) that have probabilities greater than $50\%$ of being unbound from the Galaxy. Of these, only three have a nearly $99\%$ probabilities of being unbound. On the whole HVSs subsample, there is 12 sources reported by other surveys. We study the spatial distribution of angular positions and angular separation of HVSs. We find the unbound HVSs are spatially anisotropic that is most significant in the Galactic longitude at more than $3\sigma$ level, and lower unbound probability HVSs  are systematically more isotropic. The spatial distribution can reflect the origin of HVSs and we discuss the possible origin link with the anisotropy.
\end{abstract}


\section{Introduction}

\par Hypervelocity stars (HVSs) are unbound stars escaping the gravitational potential of the Galaxy. \cite{Hills88} predicted their existence as an interaction consequence of a binary star system and the massive black hole (MBH) in the Galactic Center (GC).  In 2005, \cite{Brown05} discovered a late B-type main-sequence star having heliocentric radial velocity of $853\pm12$ km~s$^{-1}$, which is significantly in excess of the Galactic escape speed at that location.
Following the discovery of the first HVS by \cite{Brown05},  a large number of HVSs candidates have been reported \citep[e.g.,][]{Hirsch05, Edelmann05, Brown06, Brown09, Brown12, Brown14, Li12, Li18, Zheng14, Geier15, Huang17, Du18b, Marchetti18, Bromley18, Hattori18, Boubert18, Shen18}.  HVSs can not only probe the extreme dynamics and physical processes at the GC, but also can be used as dynamical traces of integral properties of the Galaxy. Such stars can be used to probe the star formation in the GC or constrain the potential of the Milky Way \citep[e.g.][]{Gnedin05}, some studies have used the kinematics of HVSs to obtain an
estimate of the Galaxy mass \citep[e.g.][]{Smith07, Piffl14}.  There have been a few studies aimed at combining their
chemical and kinematic information to get a picture of where these
stars are produced and how they achieve such high
velocities or to determine the origin of HVS
\citep[e.g.][]{Bromley09, Purcell10, Wang09, Wang13, Hawkins15, Geier15, Tauris15, Ziegerer17, Marchetti18, Du18a, Du18b, Irrgang18}.

\par Before Gaia DR2 release, 
most of the confirmed HVSs are massive early-type O- and B-type stars in the Galactic halo due to the selections bias of target surveys.   
But the HVS is not restricted to early-type 
stars, but is also observed among evolved low-mass stars such
as hot subdwarf stars \citep{Hirsch05, Geier15} and white dwarfs \citep{Vennes17, Raddi18, Shen18}.
Even some studies have been devoted to search for late-type HVSs candidates.  For example, \cite{Kollmeier09, Kollmeier10} attempted to find metal-rich, old-populations F/G type HVSs
from  sloan Digital Sky Survey (SDSS) data. However, such late-type old-population HVSs have not been detected only places a upper limit on the rate of ejection. \cite{Li12} reported the discovery of 13 F-type HVSs, located at distances ranging from 3 kpc to 10 kpc. \cite{Palladino14} found 20 low-mass G- and K-type HVS candidates from SDSS by incorporating proper motions, but none of the orbits were consistent with GC ejection. However, \cite{Ziegerer15} reanalyzed the proper motions for 14 of these and found that they are all bound to the Galaxy and the initial HVS classification was due to the flawed proper motions.  \cite{Zhong14} reported the discovery of 28 HVSs candidates which covered  a broad color range at heliocentric distance of less than 3 kpc from Large Sky Area Multi-object Fiber Spectroscopic Telescope (LAMOST) DR1. \cite{Vickers15} used more stringent proper  motions cuts to carry out a study of the runaway population of low-mass star in SDSS,  they detected  a number of high-velocity runaway stars, but their HVSs candidates were marginal detections. \cite{Zhang16} identified 29 metal-rich ([Fe/H]$>-0.8$) high velocity stars, but they did not find any candidates which have velocities in excess of the escape speed.  

\par After  Gaia DR2 release,  it has already been used to search for new HVSs that can be unbound to the Galaxy.  For example, \cite{Hattori18} reported the discovery of 30 stars with extreme velocities ($> 480$ km s$^{-1}$) in Gaia DR2, one of them is consistent with having been ejected from GC, and another has orbit that passed near the Large Magellanic Cloud. \cite{Marchetti18} found 20 HVSs have probabilities greater than 80$\%$ of being unbound from the Galaxy in the subset of stars with radial velocity measurements of Gaia DR2.  \cite{Du18b} found 24 late-type high velocity stars with stellar astrometric parameters and radial velocities from Gaia DR2 and LAMOST data.  Most of the high velocity stars are metal-poor and $\alpha$-enhanced. Of these, 6 stars belong to hypervelocity stars.   
\cite{Bromley18} investigated the nature of nearby (10-15 kpc) high-speed stars in Gaia DR2 and identified two have $100\%$ probability of unbound  in over 100 high-speed stars.   However, most previous surveys are based on the subsample of about 7 million stars with radial velocity measurements alone in Gaia DR2. As shown by \cite{Bromley18},  based on Gaia DR2 parallaxes and proper motions alone one can efficiently select relatively nearby high-speed candidates.  
Using the Galactic rest-frame tangential velocity should return a higher proportion of nearby HVSs than radial velocity \citep{Kenyon18}.
The tangential velocity need to use accurate proper motions and parallaxes with sufficiently small uncertainties. 

\par In this study, our goal is to consider the wealth of accurate proper motions and parallaxes available in Gaia DR2 \citep{Gaia18a, Gaia18b} to obtain more HVSs and study the distribution of HVS angular position on the sky and judge if there is significantly anisotropic. The paper is structured as follows. In Section 2, we briefly describe the data and HVSs sample candidate selection. In Section 3, we show the observed spatial distribution of HVSs on the sky is anisotropic. In Section 4, we discuss the possible origin of spatial anisotropy of unbound HVSs. The conclusions and summary are given in Section 5.

\section{Data and target selection}

\subsection{Gaia data}

\par The Gaia satellite is a space-based mission launched in 2013 and started science operations the following year. The second Gaia data (Gaia DR2 ) includes high-precision measurements of nearly 1.7 billion stars \citep{Gaia18a, Gaia18b}. As well as positions, the data include photometry, radial velocities, and information on astrophysical parameters and variability, for sources brighter than magnitude 21. The parallaxes, and mean proper motions for about 1.3 billion of the brightest stars are contained. Radial velocity measurements $v_{r}$ for a subset of 7,224,631 stars are included in Gaia DR2 archive \citep{Gaia16} with an effective temperature from 3550 to 6990 K within a few thousand parsec of the Sun. 
The median uncertainty for the bright sources ($G<14$ mag) is $0.04$ mas, $0.1$ mas at $G = 17$ mag, and $0.7$ mas at $G = 20$ mag for the parallax,  and $0.05, 0.2$, and $1.2$ mas\,yr$^{-1}$ for the proper motions, respectively.  More detailed description about the astrometric content of Gaia DR2 can be found in \cite{Lindegren18}.

In this study, we first select stars  with parallax uncertainties smaller than $20\%$.   In order to assure the quality of the reported radial velocity,  we keep those sources that have 
$rv\_nb\_transits >5$ in Gaia DR2 archive, indicating that $v_{r}$ measurements were taken at a minimum of six distinct epochs.    
We do not adopt other cut on the astrometric solution.  While a recent paper by \cite{Marchetti18}
adopted more conservative criteria for the quality of the Gaia astrometric solution.  As mentioned by \cite{Hattori18}, the conservative cut on the quality of astrometric solution might potentially discard a lot of interesting candidate stars with small formal errors on parallax and proper motion.  In addition, due to different Galactic potential model and  different estimation method about probability  of being unbound, the HVSs sample could be different.

\subsection{Distance derivation from Gaia parallaxes}

\par According to the parallaxes and proper motions,  distance and velocities could be inferred using those data.  This is important task to derive distance and velocities, especially when parallaxes are involved because the effects of the observational errors on the parallaxes and the proper motions  can lead to potentially strong biases \citep{Luri18}.  
\cite{Butkevich17} confirmed that due to various instrumental effects of Gaia satellite, in particular, to a certain kind of basic-angle variations, these can bias the parallax zero point of an astrometric solution derived from observations. 
From the quasars and validation solutions, \citet{Lindegren18} estimated that systematics in the parallaxes depending on position, magnitude, and color are generally below 0.1 mas, but global parallax zero-point of Gaia observations is:  $\varpi_{\rm zp}=-0.029$ mas.   Thus, it is necessary to subtract parallax zero-point ($\varpi_{\rm zp}$) when parallax is used to calculate astrophysical quantities.   

Despite the simple relation between the parallax and distance, inversion of the parallax to obtain distance is only appropriate when there are no measurement errors.
While the significance of the parallax detections in the $5\sigma$ sample is high, parallax errors are not negligible. The parallax error distribution of Gaia DR2 sources is well approximated by a Gaussian  with a tail extending to negative values \citep{Lindegren18}.  When deriving the distance by inverting the Gaia DR2 parallax, even small values allowed by the uncertainties may be unrealistic.
By comparison of the heliocentric distance derived by inverting the Gaia DR2 parallax with derived by \cite{Bailer18} using a geometrical distance prior for our sample, 
we found that the distance derived by inverting the parallax is relative precise just for nearby (less than  2 kpc) sample stars.
When the parallax error distribution is narrow compared to the measured parallax, the extending tail of the distribution are negligible especially for the nearby stars. So some studies selected only sources that have small relative parallax error \citep[e.g.][]{Hattori18} to give a straightforward estimate of distance  from the inverse of the parallax. A more general method,  Bayesian approach, incorporates prior information about source location \citep{Bailer15, Astra16, Luri18}.

In this study, in order to derive relative accurate distance estimation for distant sample stars, we use the full Bayesian approach to infer distances \citep{Luri18}. We use the exponentially decreasing space density prior in distance $d$ with a most probable source location at $2L$ \citep{Bailer18}:
\begin{equation*}
P(d\ |\ L) \propto d^2 \exp (-d/L)
\end{equation*}
Here, the length scale $L$ depending on the sky location relative to the Galaxy \citep{Bailer18} and we assume uniform priors on $v_{ra}, v_{dec}, v_r$. So we can express the posterior distribution:
\begin{equation*}
P(\bm{\theta}\ |\ \bm{x}) \propto \exp [-\frac{1}{2} (\bm{x} - \bm{m(\theta)})^\mathrm{T} C_x^{-1} (\bm{x} - \bm{m(\theta)})]\ P(d\ |\ L)
\end{equation*}
where $\bm{\theta} = (d,\ v_{\alpha},\ v_{\delta},\ v_r)^\mathrm{T}$, $\bm{x} = (\varpi,\ \mu_{\alpha^*},\ \mu_{\delta},\ rv)^\mathrm{T}$, $\bm{m} = (1/d,\ v_{\alpha}/kd, \ v_{\delta}/kd,\ v_r)^\mathrm{T}$, $k$ = 4.74 and $C_x$ is covariance matrix:
$$
\begin{pmatrix}
\sigma_{\varpi}^{2} & \rho_\varpi^{\mu_{\alpha^{*}}} \sigma_{\varpi} \sigma_{\mu_{\alpha^*}}  & \rho_\varpi^{\mu_{\delta}} \sigma_{\varpi} \sigma_{\mu_{\delta}} & \rho_\varpi^{rv} \sigma_{\varpi} \sigma_{rv} \\
\rho_\varpi^{\mu_{\alpha^{*}}} \sigma_{\varpi} \sigma_{\mu_{\alpha^{*}}} & \sigma_{\mu_{\alpha^{*}}}^{2} & \rho_{\mu_{\alpha^{*}}}^{\mu_{\delta}} \sigma_{\mu_{\alpha^{*}}}\sigma_{\mu_{\delta}} & \rho_{\mu_{\alpha^{*}}}^{rv} \sigma_{\mu_{\alpha^{*}}} \sigma_{rv} \\
\rho_\varpi^{\mu_{\delta}} \sigma_{\varpi} \sigma_{\mu_{\delta}} & \rho_{\mu_{\alpha^{*}}}^{\mu_{\delta}} \sigma_{\mu_{\alpha^{*}}}\sigma_{\mu_{\delta}} & \sigma_{\mu_{\delta}}^{2} & \rho_{\mu_{\delta}}^{rv} \sigma_{\mu_{\delta}} \sigma_{rv} \\
\rho_\varpi^{rv} \sigma_{\varpi} \sigma_{rv} & \rho_{\mu_{\alpha^{*}}}^{rv} \sigma_{\mu_{\alpha^{*}}} \sigma_{rv} & \rho_{\mu_{\delta}}^{rv} \sigma_{\mu_{\delta}} \sigma_{rv} & \sigma_{rv}^{2}
\end{pmatrix}
$$
where $\rho_i^j$ denotes the correlation coefficient between $i$ and $j$, $\sigma_k$ denotes the standard deviation of $k$ and the correlation coefficient $\rho_i^{rv}=0$, $i=\varpi, \mu_\alpha^{*}, \mu_{\delta}$.

\subsection{Coordinate Systems and Velocity Computation}
\par We then transform the Galactic coordinates $(l, b)$ and heliocentric distance for the stars into a Cartesian Galactocentric coordinate system $(X, Y, Z)$, and derive the projected distance from the Galactic center using coordinate transformations \citep{Bond10}:

\begin{equation}
\begin{split}
&X = R_{\odot} - d \cos(b) \cos(l)\\
&Y = - d \cos(b) \sin(l)\\
&Z = d \sin(b) + z_{\odot}\\
\end{split}
\end{equation}

Here, we adopt the distance from the Sun to Galactic center $R_{\odot} = 8.2$ kpc \citep{Bland16}, and the Sun has an offset from the local disk  $z_{\odot}$ = 25 pc \citep{Juric08}, $d$ is distance from the star to the Sun, and ($l$, $b$ ) are the Galactic longitude and latitude.  We adopt a Local Standard of Rest velocity $V_{{\rm LSR}} = 232.8\ {\rm km~s^{-1}}$ in the direction of rotation \citep{McMillan17}, and the solar peculiar motion ($V_X^{\odot,{\rm pec}}, V_Y^{\odot,{\rm pec}},
V_Z^{\odot,{\rm pec}}$) = ($10.0 \rm ~km\ s^{-1}, 11.0~km\ s^{-1}, 7.0~km\ s^{-1})$ \citep{Tian15, Bland16}.
We calculate each star's Galactic space-velocity components from its tangential velocities, distance, and radial velocity \citep{Johnson87}. 
 
\begin{align}
 &V_X = V_X^{{\rm obs}} + V_X^{\odot,{\rm pec}} \nonumber \\
 &V_Y = V_Y^{{\rm obs}} + V_Y^{\odot,{\rm pec}} + V_{\mathrm{LSR}}\\
 &V_Z = V_Z^{{\rm obs}} + V_Z^{\odot,{\rm pec}} \nonumber 
\end{align}
We can use 6D phase space information to derive the total velocities  $V_\mathrm{gc}$ in the Galactic rest frame and 5D information to derive the  total tangential velocity $V_\mathrm{tan}$.
$V_\mathrm{tan}$ is the Galactic rest frame tangential velocity corrected for  the solar motion, is given by:
\begin{equation}
\vec{V}_\mathrm{tan}= \vec{v}_\mathrm{tan} +\vec{V}^{\odot}-\left(\vec{V}^{\odot} \cdot \hat{r}\right) \hat{r}
\end{equation}
where $\vec{v}_\mathrm{tan} = (v_{\alpha},\ v_{\delta})$, $\vec{V}^{\odot}$ is the Sun's velocity in the Galaxy's rest frame, and $\hat{r}$ is the unit vector in the direction of the Sun in that frame.

\subsection{HVSs sample Candidate Selection}

\par The escape speed $V_{\mathrm{esc}}$ at different Galactocentric distance can be derived by adopting a Galaxy potential model which is provided by \cite{McMillan17}. Their model includes four components: the cold gas discs near the Galactic plane, the thin and thick stellar discs, a bulge and a dark-matter halo. The Galactic potential $\Phi$ is sum of potential of each component. We define unbound stars as ones which can reach the point with gravitational potential $\Phi_\mathrm{max}$ and the local escape speed $V_{\mathrm{esc}}$ is given by:
\begin{equation}
V_{\mathrm{esc}}=\sqrt{2[\Phi_\mathrm{max}-\Phi(r_\mathrm{gc})]}  
\end{equation}
where $\Phi(r)$ is the gravitational potential at $r$ and $r_\mathrm{gc}$ is Galactocentric distance. We use the local escape speed $521 \rm km~s^{-1}$ \citep{Williams17} to constrain the escape curve of \cite{McMillan17}.

\par We first identified about 6,000 candidate stars with $V_\mathrm{tan}>0.75 V_\mathrm{esc}$ or $V_\mathrm{gc}>0.75 V_\mathrm{esc}$. For each star, we use Markov chain Monte Carlo (MCMC) sampler EMCEE \citep{Goodman10, Foreman-Mackey13} to estimate error of these stars. We run each chain using 20 walkers and 1000 steps (including 500 burn-in steps), for a total 10000 random samples drawn from the posterior distribution $P(\bm{{\theta}}| \ \bm{x})$. We remove stars with $\sigma_V/V > 0.3$ and get 4565 candidate stars with $V_\mathrm{tan}>0.75 V_\mathrm{esc}$, 441 stars with $V_\mathrm{gc}>0.75 V_\mathrm{esc}$. For each star, we compute the probability $P_\mathrm{ub}$ of being unbound from the Galaxy as the fraction of Monte Carlo realization which result in an unbound orbit has $V_\mathrm{gc}\geq V_\mathrm{esc}$ at that given position.

\par
\begin{figure}[b]
	\centering
	\includegraphics[width=1.0\hsize]{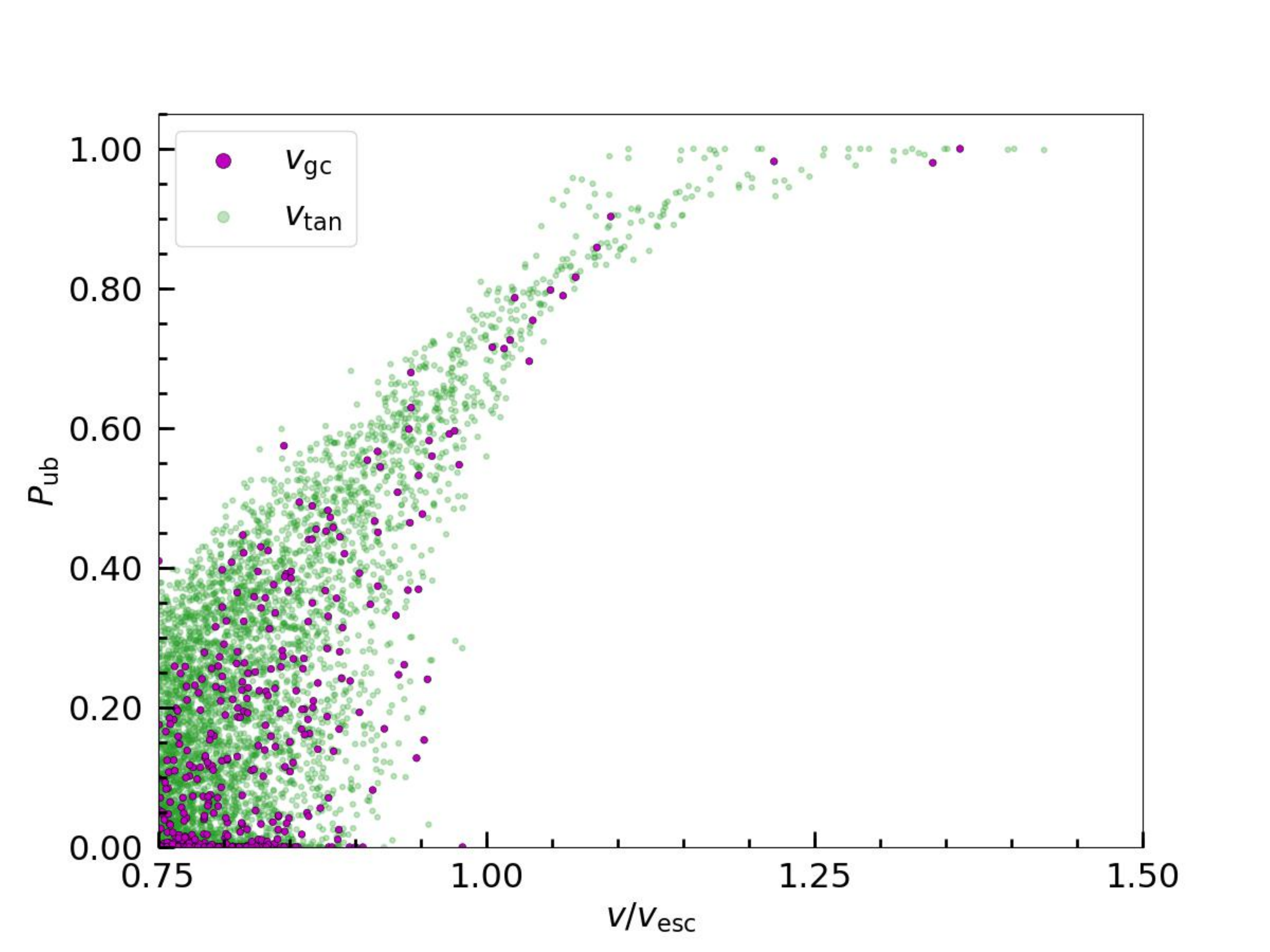}
	\caption{The probability of unbound $P_{\mathrm{ub}}$ as a function of velocity in the Galactic rest-frame $V_{\mathrm{gc}}$ for stars with $V>0.75 V_\mathrm{esc}$. The green dots represent tangential velocities from proper motion of 5D information and the purple dots represent the total velocities from radial velocity and proper motion of 6D information. Only stars with $V>0.75 V_\mathrm{esc}$ are considered in this study.}
	\label{figure1} 
\end{figure}

\begin{figure}
	\centering
\includegraphics[width=1.0\hsize]{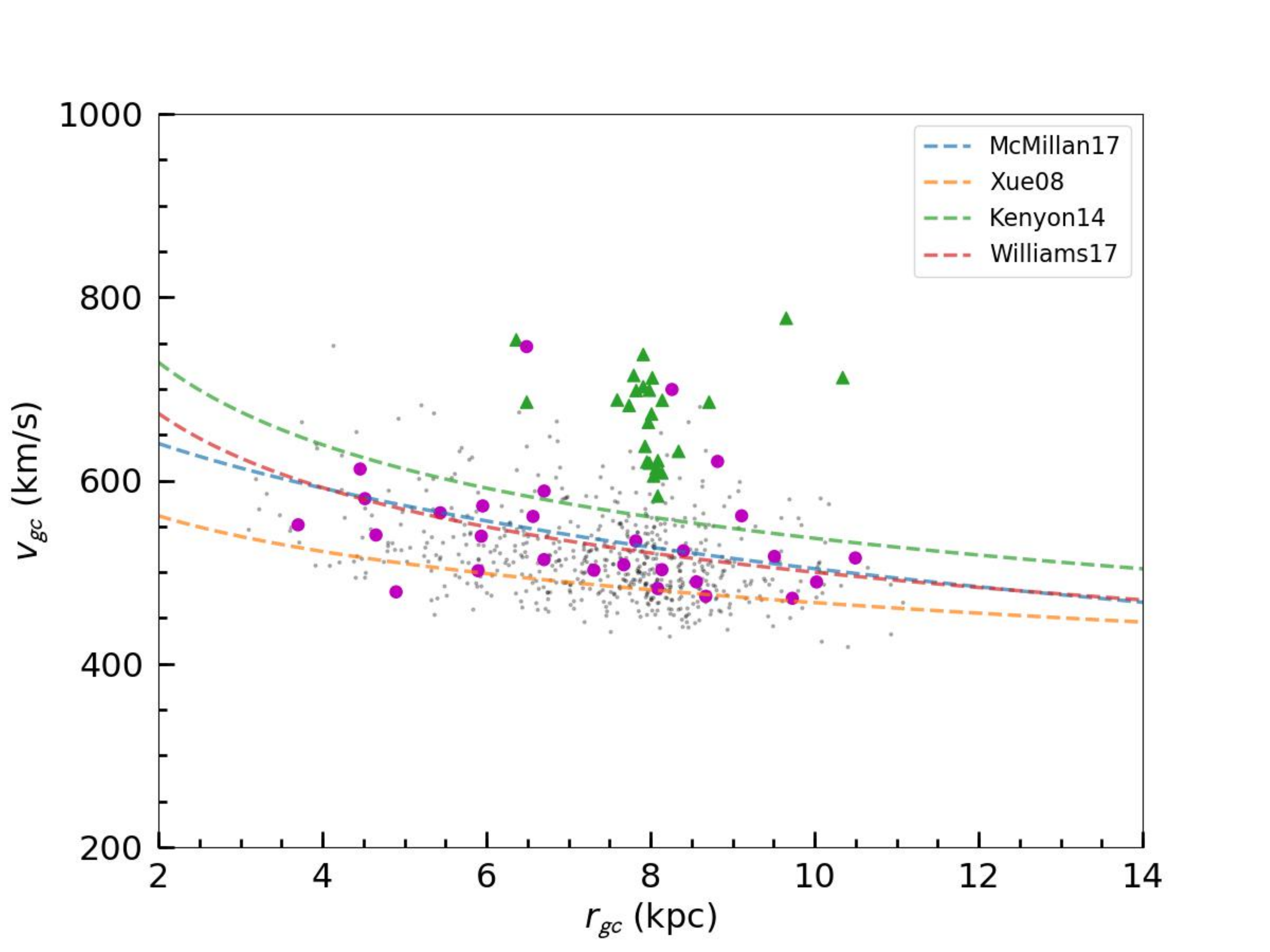}
\caption{Total velocity in the Galactic rest-frame $v_\mathrm{gc}$ as a function of Galactocentric distance $r_\mathrm{gc}$ for stars with $P_\mathrm{ub}>0.5$ (black dots) and with $P_\mathrm{ub}>0.99$ (green triangles) from 5D information and stars with $P_\mathrm{ub}>0.5$ (purple circles) from 6D information. Four different colors dashed curves are median escape speeds from different Galaxy potential models \citep{Xue08, Kenyon14, McMillan17, Williams17}.}
\label{figure2}     
\end{figure}

 Figure \ref{figure1} shows the probability of unbound $P_\mathrm{ub}$ as a function of velocity in the Galactic rest-frame $V_{\mathrm{gc}}$ for stars with $V>0.75 V_\mathrm{esc}$. The orange dots represent tangential velocity from 4565 candidates having proper motion in 5D information, and the blue dots represent the total velocity from 441 candidates having radial velocity and proper motion in 6D information. Next we identify HVSs that are potentially unbound to the Galaxy. There are 28 stars with the probability of unbound $P_\mathrm{ub}>0.99$ (red star) from 5D information and 28 stars with $P_\mathrm{ub}>0.5$ (red hexagon) from 6D information, which are shown in different signal of Figure \ref{figure2}.

In this figure, we give the velocity in the Galactic rest-frame $V_\mathrm{gc}$ as a function of Galactocentric distance $r_\mathrm{gc}$ for stars with $P_\mathrm{ub}>0.5$ (black dot) and with $P_\mathrm{ub}>0.99$ (red star) from 5D information and with $P_\mathrm{ub}>0.5$ (red hexagon) from 6D information, also present the escape speed curve from four Galactic potential models \citep{Xue08, Kenyon14, McMillan17, Williams17}, the difference of curves illustrates the some uncertainties between these models. In the appendix, Table 2 provides Gaia identifiers, position, parallaxes, proper motions, distances and the Galaxy's velocity of HVSs with the probability of unbound $P_\mathrm{ub}>0.99$ from Gaia DR2 5D information and Table 3 provides the same parameters (the last column is the the probability of unbound) with $P_\mathrm{ub}>0.5$ from Gaia DR2 6D information. In the first column, the superscript `B', `H', `M', and `S' indicates that a source is also listed in \citet{Bromley18}, \citet{Hattori18}, \citet{Marchetti18}, and \citet{Scholz18}, respectively.

\par \citet{Bromley18} investigate the probability of stars not being bound using Gaia DR2 data. We find that 5 6D-unbound sources are also listed in their research. These stars have very high $P_\mathrm{ub}$, and the little difference of $P_\mathrm{ub}$ may be due to the potential model. \citet{Hattori18} trace the orbits of stars with extremely large velocities back in time and discuss the origin of these unbound stars. A `HRS' and a `OUT' candidates are also listed in their research, but the origins are undetermined in their study. \citet{Marchetti18} search for unbound stars with Gaia DR2, and discuss their origin. We find that a `OUT' and 3 `HRS' candidates are also listed in their research, and the conclusion about the origin of stars is consistent with theirs. \citet{Scholz18} study stars with high Galactic rest frame tangential velocities, and we find that 5 5D-unbound sources are also listed in their research.

\begin{figure}
	\centering
\includegraphics[width=1.0\hsize]{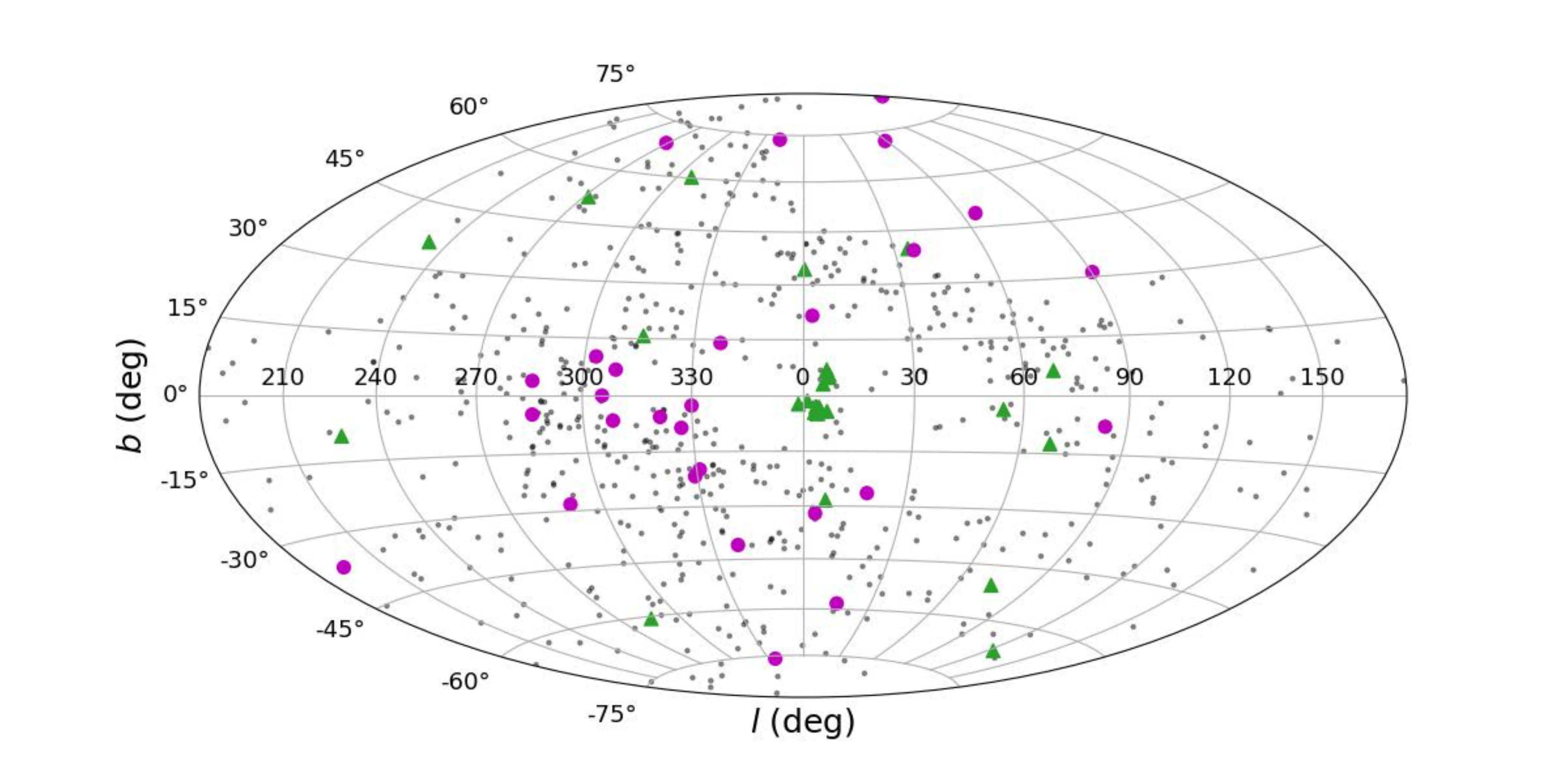}
\caption{The spatial distribution of stars in the Galactic coordinate with $P_\mathrm{ub}>0.5$ (black dots) and with $P_\mathrm{ub}>0.99$ (green triangles) from 5D information and with $P_\mathrm{ub}>0.5$ (purple circles) from 6D information. }
\label{figure3}     
\end{figure}

\begin{figure*}
	\centering
\includegraphics[width=0.9\hsize]{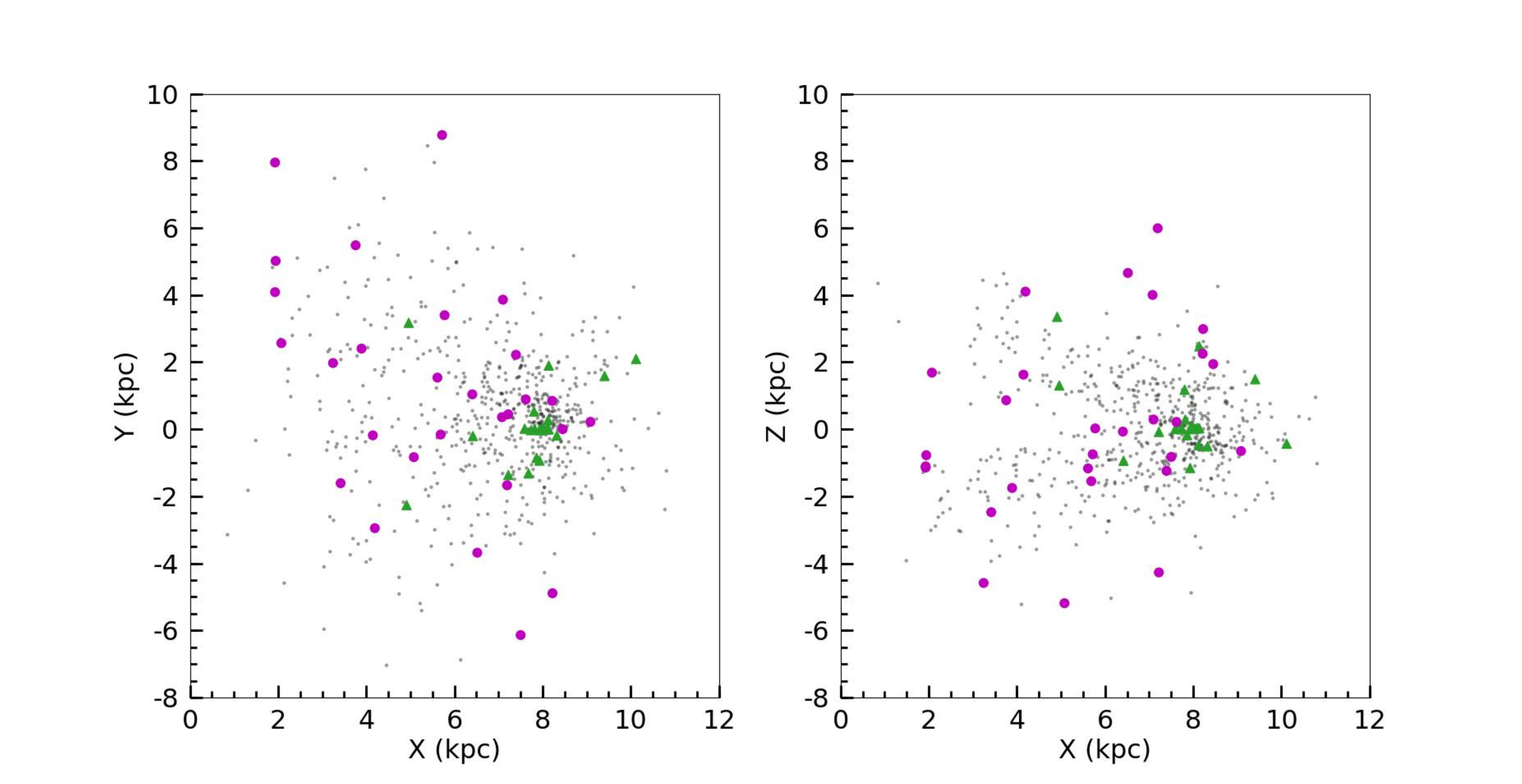}
\caption{The spatial distribution of stars  in Cartesian Galactocentric coordinate system with $P_\mathrm{ub}>0.5$ (black dots) and with $P_\mathrm{ub}>0.99$ (green triangles) from 5D information and with $P_\mathrm{ub}>0.5$ (purple circles) from 6D information.}
\label{figure4}     
\end{figure*}

\begin{figure}
	\centering
\includegraphics[width=1.0\hsize]{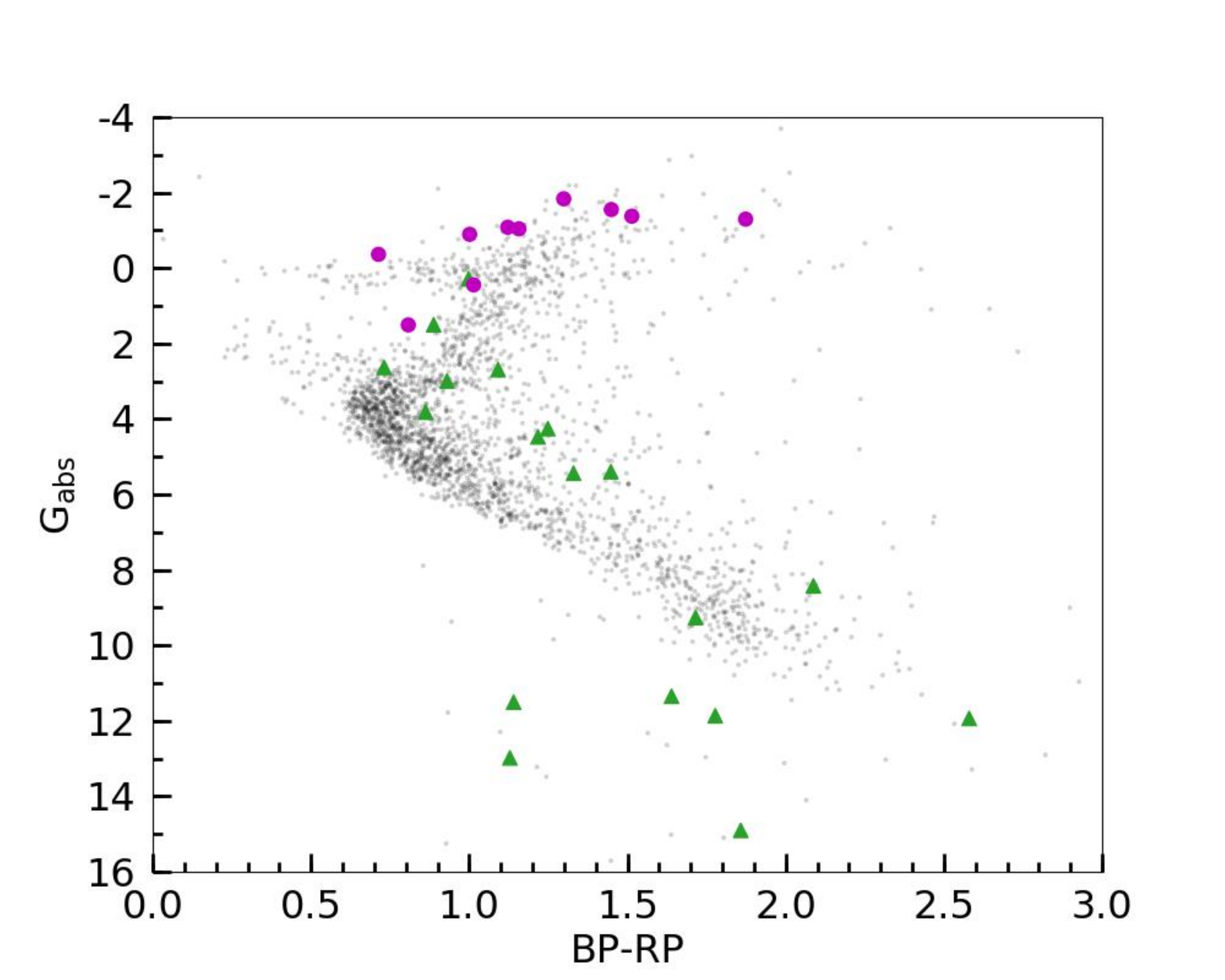}
\caption{The HR diagram for the unbound HVSs with $P_\mathrm{ub}>0.99$ (green triangles) from 5D information and HVSs with $P_\mathrm{ub}>0.5$ (purple circles) from 6D information and all stars with $V_\mathrm{gc}>0.75 V_\mathrm{esc}$ (black dots).}
\label{figure5}     
\end{figure}

\section{Spatial distribution of HVS}

\par The spatial distributions of HVSs on the sky can place useful constraints on their origin. For example, the massive black hole (MBH) in the Galactic center can eject HVSs in all directions. Using a few late B-type unbound HVSs, \cite{Brown09, Brown12, Brown14} demonstrated that unbound HVSs have a significantly anisotropic on the sky, while lower velocity bound HVSs have a more isotropic distribution. We begin to study the trend by adopting 56 nearby late-type HVSs. Figure \ref{figure3} plots the angular position of these HVSs in the Galactic coordinate. HVSs with $P_\mathrm{ub}>0.99$ (which we call 5D-unbound) are marked with green triangles and possible unbound HVSs with $P_\mathrm{ub}>0.5$ (which we call 5D-possible-ub) are marked with black dots from 5D information and possible unbound HVSs with $P_\mathrm{ub}>0.5$ (which we call 6D-unbound) from full 6D information are marked with purple circles. Figure \ref{figure4} also shows the spatial distribution of stars in Cartesian Galactocentric coordinate system.  We can notice that most of unbound HVSs in our sample are presently locate at nearby the Galactic plane, particularly a few unbound HVSs are significantly clustered together around the solar neighborhood.   

\par \citet{Green19} present a three-dimensional map of dust reddening, based on Gaia parallaxes and stellar photometry from Pan-STARRS 1 and 2MASS. We use the `bayestar2019' version of the 3D dust map and apply the median reddening for each stars. The optical extinction can be derive from the color excess:
\begin{equation}
A_V = 3.1\ E(B-V).
\end{equation}
Then we convert it to the Gaia passbands \citep{Bromley18}:
\begin{equation}
A_G \approx 0.848 A_V.
\end{equation}
Figure \ref{figure5} shows the HR diagram for 5D-unbound (green triangles) and 6D-unbound (purple circles) HVSs and all stars with $V_\mathrm{gc}>0.75 V_\mathrm{esc}$ (black dots). The x-axis represents the color index in the Gaia Blue Pass (BP) and Red Pass (RP) bands BP-RP, and the y-axis gives the absolute magnitude in the Gaia $G$ band. We can see that some unbound HVSs are giants and also a few dwarfs, and these HVSs are late type stars.

\par
\begin{figure}
	\centering
	\includegraphics[width=1.0\hsize]{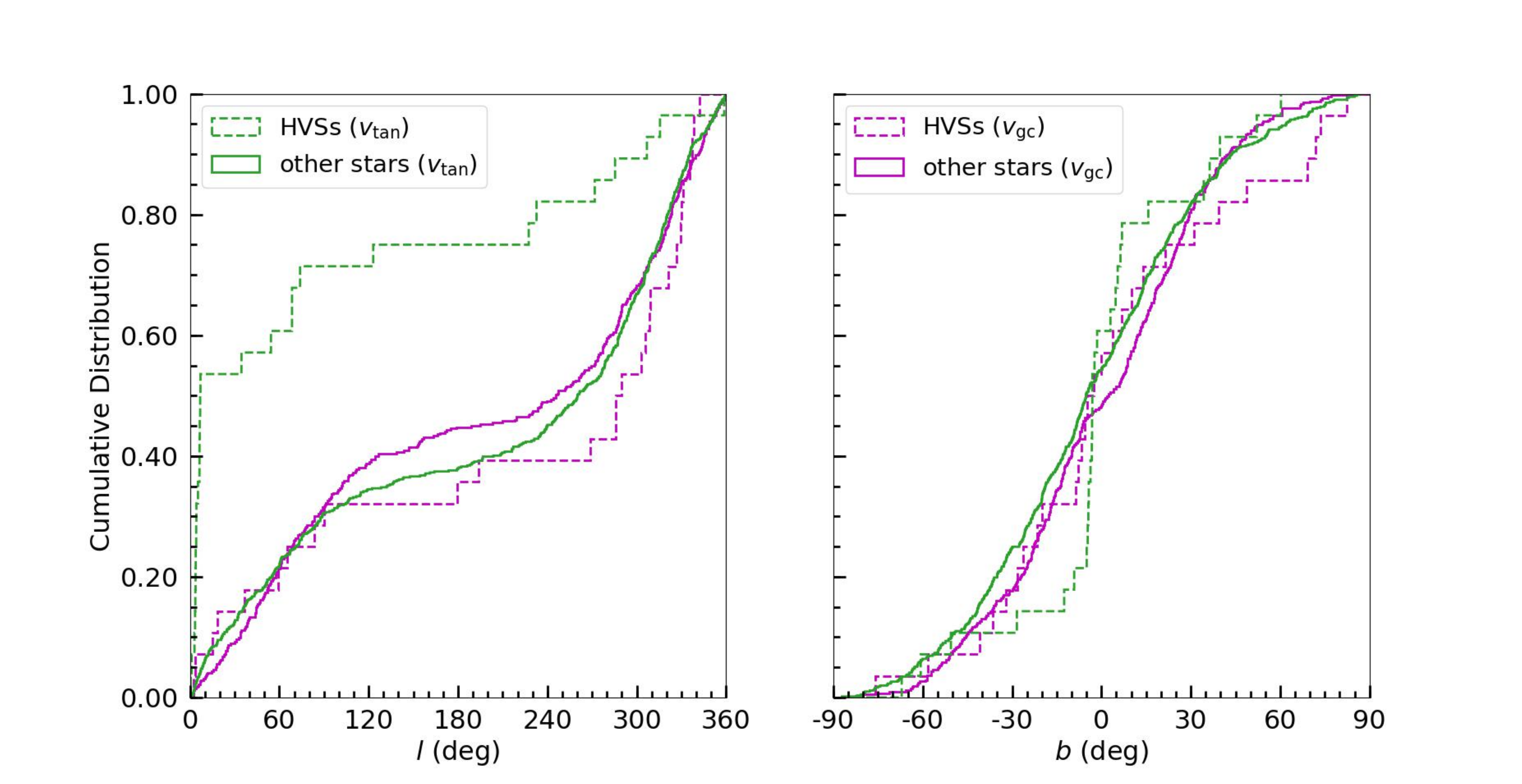}
	\caption{ Cumulative distribution of the Galactic longitude $l$ and $b$ for unbound HVSs and lower unbound probabilities HVSs. 
		The purple dashed line represent 28 lower unbound probabilities HVSs with $P_\mathrm{ub}>0.5$ and purple solid line represents other 
		stars with $P_\mathrm{ub}<0.5$ and $V>0.75 V_\mathrm{esc}$ from 6D information. The green dashed line represent 
		28 unbound HVSs with $P_\mathrm{ub}>0.99$ and the green solid line is other 619 lower velocity HVSs with $P_\mathrm{ub}>0.5$ 
		from 5D information.}
	\label{figure6}     
\end{figure}

 In order to test if the spatial distributions of unbound HVSs and lower unbound probability HVSs differ significantly, we use Kolmogorov-Smirnov (K-S) tests \citep[See also][]{Brown09, Brown12} which has the advantage of making no assumption about the distribution of data. Figure \ref{figure6} gives the cumulative distribution of the Galactic longitude $l$ and $b$ for 6D-unbound (purple line) and 5D-unbound (green line) HVSs. Firstly, K-S test derive isotropic distribution in longitudes and latitudes of 28 6D-unbound (purple line). Then for 28 5D-unbound, K-S tests give the likelihoods of $1.98\times10^{-6}$ for the unbound HVSs in same longitude distributions as the lower unbound probability. We derive the significance of 3.9$\sigma$ by taking 10,000 random draws of 5D-possible-unb HVSs. The likelihoods is 0.017 and significance is 2.37$\sigma$ for the Galactic latitude. Thus the distribution of unbound HVSs appears anisotropic in the Galactic longitude at more than 3$\sigma$ level, which verifies the results provided in \cite{Brown09, Brown12}.     \cite{de19} use the subsample with the lowest uncertainties to show 
the spatial distribution of hypervelocity
star candidates is anisotropic, but they consider the origin of such an
anisotropy is probably the result of observational biases
and selection effects.

\par
\begin{figure}
	\centering
	\includegraphics[width=0.9\hsize]{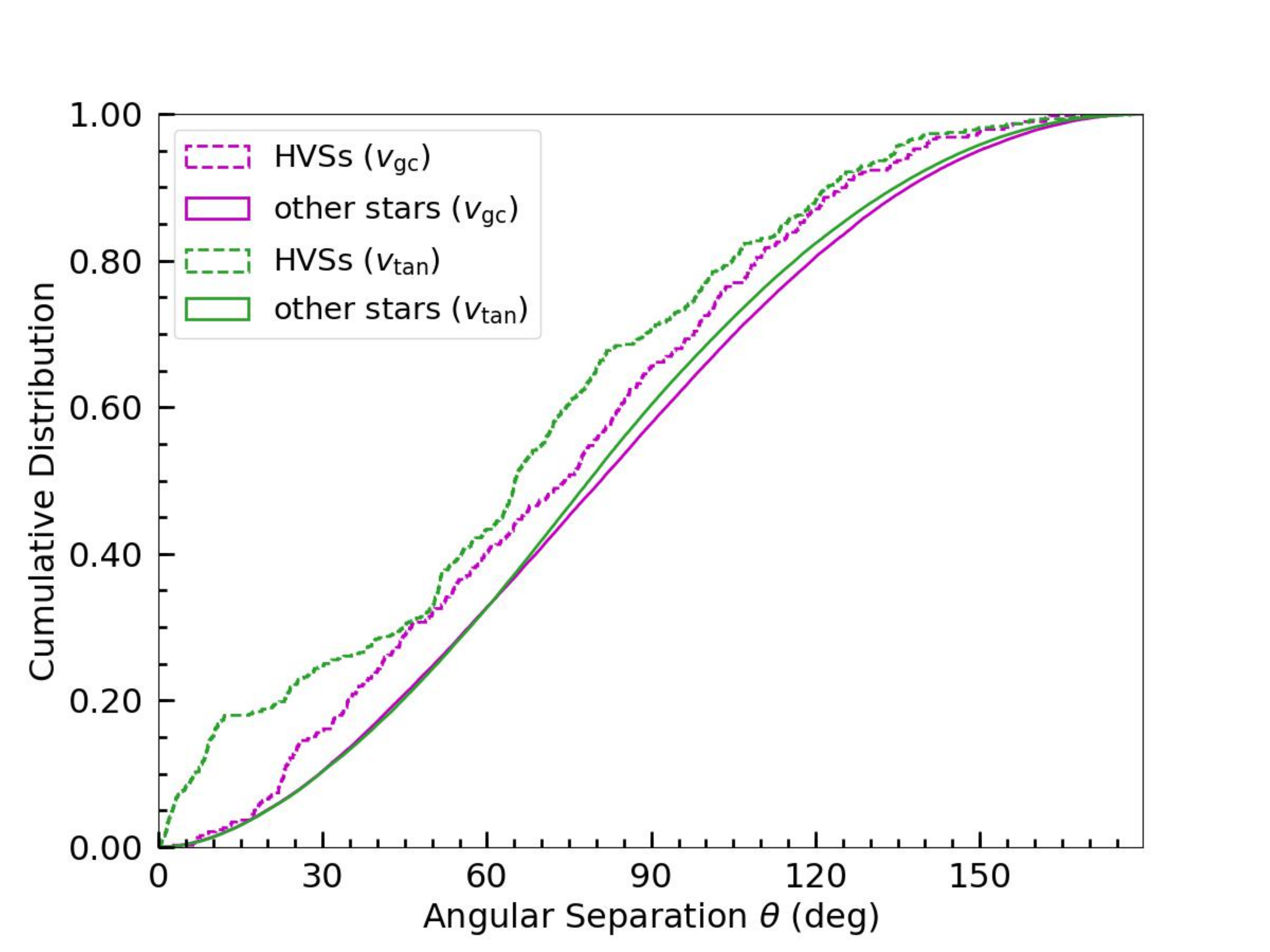}
	\caption{Cumulative distribution of angular separations for unbound HVSs  from 5D information and lower unbound probabilities HVSs from 6D information. The colors show the same as Figure 6.}
	\label{figure7}     
\end{figure}

 Similarly, we also consider the anisotropy in the distribution of angular separations of unbound HVSs compared to the lower unbound probability HVSs. Figure \ref{figure7} plots the cumulative distribution of the angular separations, $\theta$, of 6D-unbound (purple line) and 5D-unbound (green line) HVSs. Calculating the angular separations for all unique pairs of stars, a K-S test derives that lower unbound probability HVSs have a more isotropic distribution of angular separations. But for 5D-unbound HVSs, the K-S test gives a $1.34\times10^{-16}$ likelihood that are drawn from the same distribution of angular separations as the lower unbound probability HVSs and the significance is $3.3\sigma$. So the angular separation distribution of unbound HVSs also shows anisotropic. Therefore, the new HVSs discoveries that also include a few candidates in the southern sky of Gaia DR2 prove the previous claim of unbound HVS spatial anisotropy and lower velocity stars spatial isotropy distribution \citep[See][]{Brown09,Brown12}.

\section{Possible origin of the spatial anisotropy}

\par We can derive the origin of stars from the position of these star crossing the disk \citep{Marchetti18}. The minimum value of the distance from the Galactic center to the crossing point is called $R_{\mathrm{min}}$. For each 6D-unbound HVS, we use the MCMC realizations discussed above and integrate each orbit back in a total time of 5 Gyr. Then we define the probability that a star is derived from the Galactic center $P_{\mathrm{gc}}$ ($R_{\mathrm{min}} < 1$ kpc) and from the Galaxy $P_{\mathrm{MW}}$ ($R_{\mathrm{min}} < 25$ kpc, \citet{Xu15}). The classified criteria of stars' origin are shown in Table 1. We find 3 hypervelocity stars (HS) that could originate in the Galactic center, 12 stars with extragalactic origin (OUT) and 13 hyper-runaway star (HRS) candidates \citep[see also][]{Du18b}.

\begin{center}
	\begin{longtable}{cccc}
		\caption{The probability of stars are used as the classified criteria}\\
		\hline
		\hline
		Class & $P_{\mathrm{gc}}$ &   $P_{\mathrm{MW}}$ \\
		\hline
		$\ \ $HS candidates & $>0.16$ & - 	 \\
		OUT	candidates& $<0.16$ & $<0.5$ \\
		HRS	candidates & $<0.16$ & $>0.5$ \\
		\hline
	\end{longtable}
	\label{class}
\end{center}

\par A number of models have been proposed to explain the spatial anisotropy
of HVSs. Each model predicts different spatial distributions of HVSs. 
One model is the origin from the interaction between a single star and a hypothetic binary MBH \citep{Yu03, Sesana06,Sesana07,Merritt06}, the interaction between a globular cluster with  a binary MBH in the GC \citep[e.g.,][]{Capuzzo15, Fragione16}.  
The binary black hole preferentially ejects HVSs from its orbital plane.  Thus the expected
signature of a binary black hole in-spiral event
is a ring or shell of HVSs around the sky \citep{Gualandris05, Levin06, Sesana06, Sesana08}, the resulting HVS distribution in this scenario may be isotropic.  \cite{Lu10} and \cite{Zhang10, Zhang13} propose that the
HVSs anisotropy reflects the anisotropic distribution of stars in
the Galactic center. If HVSs are ejected by the central MBH, then
the direction of ejection corresponds to the direction that their
progenitors encounter the MBH. 
In addition, the Galactic center contains many structures such as the molecular gas circumnuclear disk and ionized northern arm and stellar disk.  Dynamical interactions between stellar disk and gas disk may scatter stars toward MBH to formate the S-stars and the HVSs \citep{Perets09, Perets12}. Thus the Galactic center may provide a source for the observed anisotropy of HVSs from the GC.  

\par
But as shown in \cite{Du18b},  most of nearby late-type HVSs could not origin from the Galactic center and many nearby HVSs are runaway stars candidates which originate from the disk of the Galaxy.  Some previous surveys near the Galactic disk identify unbound stars ejected from the disk \citep{Heber08, Przybilla08, Tillich09, Irrgang10, Li18} as runaway stars mechanisms \citep{Portegies00,Gvaramadze09,Bromley09,Wang13,Wang18}. The ejected star was boosted
by Galactic rotation to overcome the Galactic escape velocity.
\cite{Bromley09} compared the simulated runaways with observations and derived that simulated runaways with radial velocities exceeding 400 km s$^{-1}$ are 
located at low latitude $|b|<45$ and in the direction of Galactic rotation; observed $>400$ km s$^{-1}$ runaways at R = 50 kpc
are distributed uniformly across high latitude over all Galactic longitudes \citep{Brown12}.
Thus, the simulated distribution of runaway longitudes and latitude is inconsistent with the observed distribution of HVSs.

\par An LMC origin of some HVSs  (e.g. HE 0437-5439) was pointed out because of their proximity to the LMC on the sky.
\cite{Edelmann05} suggested HE 0437-5439 was ejected from
the Large Magellanic Cloud (LMC)  because that it was too young  to 
have traveled from the GC to its present position.
Later, the LMC origin of many more HVSs was hypothesized by \cite{Boubert16} and \cite{Boubert17}, 
which was motivated by the clustering of many HVSs in the constellation Leo \cite[see reviews][]{Brown15}. 
\cite{Abadi09} propose that the HVSs anisotropy also comes
from the stellar ejecta of a tidally disrupted dwarf galaxy which is anisotropic in the sky \citep[See also][]{Teyssier09, Piffl11}. But \cite{Brown09,Brown12} consider tidal debris an unlikely explanation for the observed HVSs because no
other unbound tidal debris is observed in the same region of sky
and no dwarf galaxy in the Local Group has high velocity in the past.   
But \cite {Fritz18} and \cite{Hammer18} derive the new results  that some dwarf galaxies have firstly high velocity using Gaia DR2 proper motions.

\par
Finally, \cite{Brown12, Brown14} propose that the HVS anisotropy
could reflect the anisotropic Galactic gravitational potential. 
Because many HVSs are marginally unbound,
a non-spherical potential can naturally explain why HVSs are
found in preferred directions on the sky: 
Stars ejected along the major axis of the potential
are decelerated less than those ejected along the minor axis.
An initially isotropic distribution of marginally unbound HVSs
can thus appear anisotropic in the halo. The predicted distribution
of HVSs depends on the axis ratio and the rotation direction
of the potential.  Up to now,  there are no good constraints on
the shape and orientation of the Galactic potential, it is difficult to judge
if HVSs anisotropy origin from Galactic gravitational potential.

\section{Conclusions and summary}

\par Based on Gaia DR2 5D information of stars with parallax and proper  motion but no radial velocity, we define our HVSs sample as those stars with $V_\mathrm{tan}>0.75 V_\mathrm{esc}$ and identify 28 nearby ( less than 6 kpc) late-type HVSs
with nearly $99\%$ unbound possibility.  At the same time,  we also define the HVSs candidate with $V_\mathrm{gc}>0.75 V_\mathrm{esc}$ from Gaia DR2 6D information from parallax, proper motion and radial velocity, 28 stars have probabilities greater than $50\%$ of being unbound from the Galaxy, but only three has a nearly $99\%$ probabilities of being unbound.  
 It also indicates for these distance ($\sim10$ kpc) that is closer to the sun and to the Galactic center, tangential motion is more significant for nearby HVSs.  As predicted by \cite{Kenyon18},  radial velocities select HVSs at large distances beyond $\sim10$ kpc, tangential velocities selects the nearby fast-moving stars.  On our HVSs sample, there is 12 sources reported by other works and they are also marked in the Table. 

\par We use Kolmogorov-Smirnov (K-S) tests to determine the spatial distributions of unbound HVSs  and lower unbound probability HVSs.   Lower unbound probability HVSs are more isotropic and unbound HVSs are spatially anisotropic at more than $3\sigma$ level, particularly in the Galactic longitude and angular separation, which strengthen the results provided in \cite{Brown09, Brown12} by using more HVSs distributed in the whole sky. In addition, most of the confirmed HVSs in previous studies are distant early-type stars 
due to the selections bias of target surveys, which is different from the nearby late-type HVSs in this study, so the new identified HVSs have different spatial distributions from previously identified sources.
The observed spatial anisotropy of HVSs could be linked to their origin.  We simply discuss the possible origin of HVSs.  In the near future, we also look forward to the accurate radial velocity observation for our HVSs sample and it can provide the trajectory calculations to give a direct test of HVSs origin. We also expect the future surveys will provide a rich source of more HVSs discoveries.

\section*{Acknowledgements}

\par  We thank especially the referee for insightful comments and suggestions, which have improved the paper significantly.
This work was supported by joint funding for Astronomy by the National Natural Science Foundation of China and the Chinese Academy of Science, under Grants U1231113.  This work was also by supported by the Special funds of cooperation between the Institute and the University of the Chinese Academy of Sciences, and China Scholarship Council (CSC).  In addition, this work was supported by the National Natural Foundation of China (NSFC No.11625313 and No.11573035).
HJN acknowledges funding from NSF grant AST 16-15688. Funding for SDSS-III has been provided by the Alfred P. Sloan Foundation, the Participating Institutions, the National Science Foundation, and the U.S. Department of Energy Office of Science. 
This project was developed in part at the 2016 NYC Gaia Sprint, hosted by the Center for Computational Astrophysics at the Simons Foundation in New York City.  This work has made use of data from the European Space Agency (ESA) mission Gaia (http://www.cosmos.esa.int/gaia), processed by the Gaia Data Processing and Analysis Consortium (DPAC, http://www.cosmos.esa.int/web/gaia/dpac/consortium). Funding for DPAC has been provided by national institutions, in particular the institutions participating in the Gaia Multilateral Agreement.

\newpage
\appendix  
\renewcommand{\appendixname}{Appendix~\Alph{section}}  
\centering
\begin{center}
	\begin{longtable}{lrrrrrrr}
\caption{HVS identified from tangential velocity selection of Gaia DR2 5D information}\\
\hline
\hline
           \hspace{\fill}source-id\hspace{\fill} & \hspace{\fill}($l,\ b$)\hspace{\fill} & \hspace{\fill}$\varpi$\hspace{\fill} & \hspace{\fill}$\mu_{\alpha^*}$\hspace{\fill} & \hspace{\fill}$\mu_{\delta}$\hspace{\fill} & \hspace{\fill}$d$\hspace{\fill} & \hspace{\fill}$r_{gc}$\hspace{\fill} & \hspace{\fill}$v_{gc}$\hspace{\fill} \\
           &	\hspace{\fill}(deg)\hspace{\fill}	&	\hspace{\fill}(mas)\hspace{\fill}	&	\hspace{\fill}(mas yr$^{-1}$)\hspace{\fill}		&	\hspace{\fill}(mas yr$^{-1}$)\hspace{\fill}		&	\hspace{\fill}(kpc)\hspace{\fill}	&	\hspace{\fill}(kpc)\hspace{\fill}	&	\hspace{\fill}(km~s$^{-1}$)\hspace{\fill}	\\
\hline
 \hspace{1.5ex}6698855754225352192$^\mathrm{S}$ &    (6.57, -28.38) &  0.494 $\pm$ 0.047 &   -38.97 $\pm$ 0.07 &   -86.57 $\pm$ 0.05 &  $2.11^{+0.24}_{-0.19}$ &   $6.44^{+0.15}_{-0.18}$ &    $717^{+106}_{-85}$ \\
\hspace{1.5ex}3593446274383096448$^\mathrm{S}$ &    (271.8, 52.16) &   0.299 $\pm$ 0.04 &   -36.35 $\pm$ 0.06 &   -47.94 $\pm$ 0.04 &  $3.26^{+0.48}_{-0.35}$ &   $8.77^{+0.18}_{-0.12}$ &   $734^{+136}_{-100}$ \\
\hspace{1.5ex}4119670443586493184 &      (6.86, 4.77) &  4.211 $\pm$ 0.585 &   249.85 $\pm$ 1.79 &   216.73 $\pm$ 1.51 &  $0.28^{+0.06}_{-0.04}$ &   $7.93^{+0.04}_{-0.06}$ &     $670^{+99}_{-68}$ \\
\hspace{1.5ex}1842456376310935552 &   (68.53, -12.51) &  1.092 $\pm$ 0.175 &  -131.96 $\pm$ 0.29 &  -108.69 $\pm$ 0.31 &  $1.09^{+0.27}_{-0.19}$ &   $7.87^{+0.05}_{-0.06}$ &   $824^{+220}_{-154}$ \\
\hspace{1.5ex}4396109004117478656 &     (0.46, 34.34) &  2.215 $\pm$ 0.225 &  -359.69 $\pm$ 0.46 &  -228.42 $\pm$ 0.29 &  $0.47^{+0.05}_{-0.05}$ &   $7.81^{+0.04}_{-0.04}$ &    $728^{+102}_{-90}$ \\
\hspace{1.5ex}4912074832815572224 &  (285.44, -60.68) &  1.758 $\pm$ 0.254 &  -101.33 $\pm$ 0.39 &   -246.8 $\pm$ 0.36 &  $0.63^{+0.12}_{-0.08}$ &   $8.14^{+0.01}_{-0.01}$ &   $756^{+151}_{-101}$ \\
\hspace{1.5ex}4119759641435130624 &      (7.04, 5.46) &  4.807 $\pm$ 0.676 &   285.92 $\pm$ 1.87 &   244.62 $\pm$ 1.66 &  $0.24^{+0.06}_{-0.04}$ &   $7.96^{+0.03}_{-0.05}$ &     $663^{+98}_{-62}$ \\
\hspace{1.5ex}4065092915465740160 &     (6.43, -4.25) &  6.248 $\pm$ 0.638 &   335.49 $\pm$ 1.49 &   342.42 $\pm$ 1.44 &  $0.17^{+0.02}_{-0.02}$ &   $8.03^{+0.02}_{-0.02}$ &     $621^{+55}_{-40}$ \\
\hspace{1.5ex}2946665465655257472$^\mathrm{B}$ &   (227.47, -9.16) &  0.345 $\pm$ 0.057 &     54.42 $\pm$ 0.1 &    11.06 $\pm$ 0.13 &  $3.26^{+0.76}_{-0.54}$ &  $10.65^{+0.63}_{-0.43}$ &   $810^{+198}_{-139}$ \\
\hspace{1.5ex}4467323685050696960 &     (34.57, 39.8) &  0.177 $\pm$ 0.028 &     -2.8 $\pm$ 0.03 &   -36.58 $\pm$ 0.03 &   $5.58^{+0.9}_{-0.67}$ &   $6.39^{+0.11}_{-0.02}$ &   $819^{+155}_{-115}$ \\
\hspace{1.5ex}3841458366321558656$^\mathrm{S}$ &    (232.7, 36.34) &  0.363 $\pm$ 0.063 &     7.29 $\pm$ 0.11 &   -81.39 $\pm$ 0.11 &  $2.63^{+0.45}_{-0.32}$ &   $9.77^{+0.31}_{-0.21}$ &   $840^{+175}_{-122}$ \\
\hspace{1.5ex}2655054950237153664 &   (73.95, -50.53) &    0.65 $\pm$ 0.03 &    84.99 $\pm$ 0.05 &   -65.81 $\pm$ 0.05 &  $1.55^{+0.07}_{-0.07}$ &     $8.07^{+0.0}_{-0.0}$ &     $621^{+37}_{-33}$ \\
\hspace{1.5ex}4062943374550066560 &     (3.44, -3.05) &  9.043 $\pm$ 0.614 &   494.36 $\pm$ 1.08 &   439.52 $\pm$ 0.87 &  $0.11^{+0.01}_{-0.01}$ &   $8.09^{+0.01}_{-0.01}$ &     $589^{+28}_{-23}$ \\
\hspace{1.5ex}4121504428896960640 &       (6.38, 6.9) &  9.146 $\pm$ 0.932 &   580.45 $\pm$ 2.88 &   453.91 $\pm$ 2.37 &  $0.12^{+0.02}_{-0.01}$ &   $8.08^{+0.01}_{-0.02}$ &     $648^{+57}_{-47}$ \\
\hspace{1.5ex}2047267531138612864 &     (68.53, 6.39) &  0.714 $\pm$ 0.081 &    -0.26 $\pm$ 0.11 &    96.36 $\pm$ 0.16 &    $1.5^{+0.2}_{-0.18}$ &   $7.78^{+0.04}_{-0.04}$ &     $748^{+92}_{-80}$ \\
\hspace{1.5ex}4050754184394584192 &      (3.03, -4.6) &   3.811 $\pm$ 0.39 &   224.82 $\pm$ 0.84 &   223.44 $\pm$ 0.79 &  $0.28^{+0.03}_{-0.03}$ &   $7.92^{+0.03}_{-0.03}$ &     $656^{+50}_{-42}$ \\
\hspace{1.5ex}2525871954701579904 &  (123.07, -67.05) &  1.709 $\pm$ 0.062 &   257.34 $\pm$ 0.14 &   -162.88 $\pm$ 0.1 &  $0.59^{+0.02}_{-0.02}$ &   $8.34^{+0.01}_{-0.01}$ &     $636^{+30}_{-27}$ \\
\hspace{1.5ex}3685380427311132160 &   (306.79, 60.27) &  0.752 $\pm$ 0.063 &    26.92 $\pm$ 0.14 &  -134.33 $\pm$ 0.07 &   $1.36^{+0.12}_{-0.1}$ &   $7.91^{+0.01}_{-0.01}$ &     $761^{+73}_{-65}$ \\
\hspace{1.5ex}4050705707044344832 &     (3.87, -4.94) &   4.499 $\pm$ 0.46 &   283.27 $\pm$ 0.96 &    281.43 $\pm$ 0.9 &  $0.24^{+0.03}_{-0.02}$ &   $7.96^{+0.02}_{-0.03}$ &     $685^{+50}_{-45}$ \\
\hspace{1.5ex}4052470217026086144 &     (4.19, -4.39) &   5.779 $\pm$ 0.76 &   413.64 $\pm$ 1.27 &   385.05 $\pm$ 1.13 &   $0.2^{+0.04}_{-0.03}$ &    $8.0^{+0.03}_{-0.04}$ &    $772^{+101}_{-75}$ \\
\hspace{1.5ex}4055741088022263680 &   (358.71, -2.32) &  1.722 $\pm$ 0.264 &   119.84 $\pm$ 1.04 &   102.69 $\pm$ 0.87 &  $0.69^{+0.17}_{-0.11}$ &   $7.51^{+0.11}_{-0.17}$ &    $751^{+125}_{-84}$ \\
\hspace{1.5ex}4057367235624058240 &     (1.14, -1.46) &  5.505 $\pm$ 0.667 &   373.14 $\pm$ 1.31 &   324.91 $\pm$ 1.04 &   $0.2^{+0.04}_{-0.03}$ &    $8.0^{+0.03}_{-0.04}$ &     $711^{+83}_{-59}$ \\
\hspace{1.5ex}4062949009547937152 &     (3.59, -3.05) &   2.21 $\pm$ 0.241 &   148.13 $\pm$ 0.91 &   140.71 $\pm$ 0.87 &  $0.49^{+0.06}_{-0.05}$ &   $7.71^{+0.05}_{-0.06}$ &     $708^{+63}_{-54}$ \\
\hspace{1.5ex}4063041295654767488 &     (3.77, -3.23) &  3.779 $\pm$ 0.401 &   498.53 $\pm$ 0.78 &   445.61 $\pm$ 0.69 &  $0.29^{+0.03}_{-0.03}$ &   $7.91^{+0.03}_{-0.03}$ &  $1142^{+109}_{-101}$ \\
\hspace{1.5ex}4063270135632421888 &     (4.33, -3.13) &  4.836 $\pm$ 0.665 &   325.47 $\pm$ 1.49 &   319.65 $\pm$ 1.27 &  $0.24^{+0.06}_{-0.03}$ &   $7.96^{+0.03}_{-0.06}$ &    $751^{+126}_{-74}$ \\
\hspace{1.5ex}4068950625022563584 &      (5.38, 3.03) &  16.07 $\pm$ 1.036 &   980.72 $\pm$ 2.86 &    799.44 $\pm$ 2.4 &    $0.06^{+0.0}_{-0.0}$ &     $8.14^{+0.0}_{-0.0}$ &     $611^{+25}_{-22}$ \\
\hspace{1.5ex}1820931585123817728 &     (54.4, -3.67) &  0.602 $\pm$ 0.076 &   -82.41 $\pm$ 0.09 &  -149.53 $\pm$ 0.09 &   $1.8^{+0.31}_{-0.21}$ &    $7.3^{+0.09}_{-0.12}$ &  $1319^{+252}_{-173}$ \\
\hspace{1.5ex}6097052289696317952$^\mathrm{S}$ &    (315.6, 15.74) &  0.202 $\pm$ 0.033 &    -61.1 $\pm$ 0.05 &   -24.73 $\pm$ 0.05 &   $5.1^{+0.85}_{-0.68}$ &    $6.0^{+0.11}_{-0.04}$ &  $1407^{+265}_{-214}$ \\
\hline
\end{longtable}
\end{center}

\begin{center}
	\begin{longtable}{lrrrrrrrrr}
		\caption{HVS identified from total velocity selection of Gaia DR2 6D information}\\
		 \hline
		\hline
           \hspace{\fill}source-id\hspace{\fill} & \hspace{\fill}($l,\ b$)\hspace{\fill} & \hspace{\fill}$\varpi$\hspace{\fill} & \hspace{\fill}$\mu_{\alpha^*}$\hspace{\fill} & \hspace{\fill}$\mu_{\delta}$\hspace{\fill} & \hspace{\fill}$rv$\hspace{\fill} & \hspace{\fill}$d$\hspace{\fill} & \hspace{\fill}$r_{gc}$\hspace{\fill} & \hspace{\fill}$v_{gc}$\hspace{\fill} & \hspace{\fill}$P_\mathrm{ub}$\hspace{\fill}\\
&	\hspace{\fill}(deg)\hspace{\fill}	&	\hspace{\fill}(mas)\hspace{\fill}	&	\hspace{\fill}(mas yr$^{-1}$)\hspace{\fill}		&	\hspace{\fill}(mas yr$^{-1}$)\hspace{\fill}	&	\hspace{\fill}(km~s$^{-1}$)\hspace{\fill}	&	\hspace{\fill}(kpc)\hspace{\fill}	&	\hspace{\fill}(kpc)\hspace{\fill}	&	\hspace{\fill}(km~s$^{-1}$)\hspace{\fill}	& \\
\hline\\
\textbf{HS candidates} &&&&&&&\\
6075020928535605632 &    (303.2, 10.18) &  0.943 $\pm$ 0.107 &  -113.45 $\pm$ 0.11 &     3.83 $\pm$ 0.1 &   -55.9 $\pm$ 0.7 &   $1.16^{+0.16}_{-0.13}$ &   $7.64^{+0.06}_{-0.07}$ &    $546^{+77}_{-63}$ &             0.58 \\
6009917672524909312 &   (337.28, 14.04) &  0.138 $\pm$ 0.023 &   -20.43 $\pm$ 0.05 &   -5.98 $\pm$ 0.03 &  -145.4 $\pm$ 1.2 &   $7.52^{+1.55}_{-1.05}$ &   $3.76^{+0.36}_{-0.08}$ &   $612^{+146}_{-94}$ &             0.60 \\
5830453567907178368 &   (326.91, -8.61) &   0.126 $\pm$ 0.02 &   -12.56 $\pm$ 0.03 &  -16.59 $\pm$ 0.03 &   155.8 $\pm$ 1.1 &   $8.36^{+1.49}_{-1.23}$ &   $4.85^{+0.66}_{-0.24}$ &  $619^{+146}_{-121}$ &             0.63 \\
\vspace{0.1ex}\\
\textbf{OUT candidates} &&&&&&&\\
3905884598043829504 &   (268.95, 69.21) &  0.406 $\pm$ 0.041 &   -35.88 $\pm$ 0.08 &  -52.22 $\pm$ 0.04 &   149.7 $\pm$ 1.2 &   $2.49^{+0.28}_{-0.24}$ &   $8.59^{+0.09}_{-0.07}$ &    $522^{+83}_{-70}$ &             0.53 \\
1436193190692538496 &    (90.29, 31.18) &  0.161 $\pm$ 0.017 &    10.68 $\pm$ 0.04 &   -9.43 $\pm$ 0.04 &  -213.3 $\pm$ 0.8 &   $5.93^{+0.61}_{-0.52}$ &   $10.15^{+0.37}_{-0.3}$ &    $503^{+41}_{-35}$ &             0.55 \\
6694147709857729920 &    (3.64, -32.07) &  0.337 $\pm$ 0.048 &   -27.19 $\pm$ 0.07 &   -42.4 $\pm$ 0.05 &   -20.0 $\pm$ 0.4 &   $3.26^{+0.61}_{-0.47}$ &    $5.7^{+0.31}_{-0.37}$ &  $570^{+142}_{-108}$ &             0.55 \\
6630656964968299264 &  (330.86, -19.83) &  0.181 $\pm$ 0.035 &   -16.32 $\pm$ 0.03 &  -22.68 $\pm$ 0.04 &   148.3 $\pm$ 1.2 &    $6.13^{+1.4}_{-1.05}$ &    $4.77^{+0.26}_{-0.1}$ &  $596^{+184}_{-137}$ &             0.57 \\
6437621133019281408 &  (329.26, -21.67) &  0.308 $\pm$ 0.029 &    -36.1 $\pm$ 0.03 &  -29.25 $\pm$ 0.04 &   252.9 $\pm$ 0.9 &   $3.38^{+0.39}_{-0.31}$ &    $5.86^{+0.17}_{-0.2}$ &    $570^{+83}_{-64}$ &             0.60 \\
5862473747352575104 &   (305.49, -0.04) &  0.232 $\pm$ 0.045 &    -29.0 $\pm$ 0.05 &  -14.44 $\pm$ 0.06 &    80.3 $\pm$ 1.2 &    $4.86^{+1.4}_{-0.88}$ &   $6.71^{+0.14}_{-0.03}$ &  $616^{+212}_{-132}$ &             0.68 \\
1478837543019912064$^\mathrm{B}$ &    (59.01, 71.88) &  0.134 $\pm$ 0.019 &   -17.61 $\pm$ 0.02 &  -16.57 $\pm$ 0.03 &  -245.9 $\pm$ 1.5 &    $6.5^{+0.79}_{-0.67}$ &   $9.63^{+0.45}_{-0.35}$ &    $541^{+86}_{-72}$ &             0.70 \\
5212110596595560192$^\mathrm{H}$ &  (289.93, -28.26) &  0.372 $\pm$ 0.018 &     9.98 $\pm$ 0.04 &  -34.43 $\pm$ 0.04 &   298.2 $\pm$ 0.5 &   $2.71^{+0.13}_{-0.13}$ &    $7.82^{+0.01}_{-0.0}$ &    $540^{+21}_{-21}$ &             0.71 \\
5827258039189703808 &   (321.27, -5.71) &   0.116 $\pm$ 0.02 &   -17.34 $\pm$ 0.03 &    -9.3 $\pm$ 0.03 &    51.1 $\pm$ 3.1 &    $8.79^{+1.7}_{-1.24}$ &    $5.71^{+0.9}_{-0.41}$ &  $633^{+157}_{-115}$ &             0.72 \\
4998248468230919936 &   (335.6, -75.95) &  0.184 $\pm$ 0.036 &    16.68 $\pm$ 0.05 &   -30.9 $\pm$ 0.05 &  -139.8 $\pm$ 0.8 &   $4.76^{+0.78}_{-0.61}$ &    $8.51^{+0.3}_{-0.19}$ &   $575^{+124}_{-96}$ &             0.73 \\
6851832593704580224 &   (18.63, -26.36) &   0.168 $\pm$ 0.03 &    -9.28 $\pm$ 0.05 &  -28.68 $\pm$ 0.03 &   -11.9 $\pm$ 2.4 &   $6.37^{+1.35}_{-1.02}$ &   $4.42^{+0.28}_{-0.09}$ &  $685^{+193}_{-144}$ &             0.79 \\
4326973843264734208$^\mathrm{M}$$^,$$^\mathrm{B}$ &      (2.6, 21.53) &  0.228 $\pm$ 0.031 &   -20.55 $\pm$ 0.05 &  -33.97 $\pm$ 0.03 &  -220.4 $\pm$ 2.1 &   $4.66^{+0.78}_{-0.58}$ &    $4.24^{+0.42}_{-0.5}$ &  $667^{+141}_{-104}$ &             0.82 \\
\vspace{0.1ex}\\
\textbf{HRS candidates} &&&&&&&\\
1964858172545721472 &    (83.52, -7.85) &  0.148 $\pm$ 0.024 &    10.72 $\pm$ 0.04 &   10.55 $\pm$ 0.04 &  -151.0 $\pm$ 0.8 &   $6.71^{+1.16}_{-0.92}$ &    $9.99^{+0.71}_{-0.5}$ &    $505^{+81}_{-64}$ &             0.51 \\
4015088951907615744 &   (179.69, 82.44) &  0.482 $\pm$ 0.081 &   -62.98 $\pm$ 0.13 &   -6.65 $\pm$ 0.09 &   -54.7 $\pm$ 0.9 &    $2.12^{+0.4}_{-0.31}$ &   $8.74^{+0.15}_{-0.11}$ &   $524^{+113}_{-84}$ &             0.54 \\
6601618897232424064 &   (14.97, -58.22) &  0.134 $\pm$ 0.025 &    18.82 $\pm$ 0.04 &   -11.0 $\pm$ 0.04 &  -137.5 $\pm$ 2.2 &   $6.47^{+1.02}_{-0.73}$ &   $7.41^{+0.37}_{-0.19}$ &    $534^{+99}_{-70}$ &             0.56 \\
5353013183518841728 &     (285.9, 3.74) &  0.239 $\pm$ 0.043 &    -9.21 $\pm$ 0.07 &   25.45 $\pm$ 0.06 &   275.1 $\pm$ 0.8 &    $4.49^{+1.0}_{-0.68}$ &    $8.2^{+0.33}_{-0.16}$ &   $542^{+129}_{-87}$ &             0.57 \\
3741464316420909952 &   (342.32, 73.51) &  0.209 $\pm$ 0.036 &    -29.0 $\pm$ 0.08 &     4.3 $\pm$ 0.04 &   -55.4 $\pm$ 2.2 &    $4.4^{+0.67}_{-0.55}$ &     $8.2^{+0.2}_{-0.13}$ &    $534^{+86}_{-69}$ &             0.59 \\
5252990984378265984 &   (285.79, -4.86) &  0.103 $\pm$ 0.018 &   -12.07 $\pm$ 0.03 &    5.35 $\pm$ 0.03 &   109.2 $\pm$ 1.4 &   $9.93^{+2.13}_{-1.38}$ &  $11.03^{+1.58}_{-0.92}$ &   $564^{+130}_{-84}$ &             0.75 \\
6456587609813249536$^\mathrm{M}$$^,$$^\mathrm{B}$ &   (338.3, -40.85) &  0.128 $\pm$ 0.019 &     13.0 $\pm$ 0.03 &  -18.26 $\pm$ 0.03 &   -15.9 $\pm$ 2.8 &    $7.48^{+1.2}_{-0.86}$ &   $6.06^{+0.44}_{-0.18}$ &   $616^{+124}_{-88}$ &             0.79 \\
4563629049534297216 &    (36.39, 39.33) &  0.131 $\pm$ 0.026 &   -23.11 $\pm$ 0.03 &   -4.04 $\pm$ 0.04 &   -41.7 $\pm$ 0.7 &   $6.97^{+1.39}_{-0.91}$ &     $6.7^{+0.51}_{-0.2}$ &   $619^{+150}_{-97}$ &             0.80 \\
6065230602133664000 &    (309.01, 6.77) &  0.132 $\pm$ 0.021 &   -21.32 $\pm$ 0.04 &   -6.07 $\pm$ 0.04 &   104.0 $\pm$ 1.1 &   $7.72^{+1.34}_{-0.98}$ &   $6.91^{+0.61}_{-0.31}$ &  $653^{+140}_{-103}$ &             0.86 \\
3252546886080448384$^\mathrm{H}$ &  (193.87, -36.61) &  0.875 $\pm$ 0.064 &    16.45 $\pm$ 0.09 &  -136.1 $\pm$ 0.06 &     1.7 $\pm$ 4.9 &   $1.16^{+0.09}_{-0.08}$ &   $9.13^{+0.07}_{-0.06}$ &    $574^{+57}_{-48}$ &             0.90 \\
5847216962695435392 &   (308.26, -6.59) &  0.087 $\pm$ 0.017 &   -15.91 $\pm$ 0.02 &   -7.34 $\pm$ 0.03 &  260.1 $\pm$ 18.2 &  $11.55^{+2.19}_{-1.89}$ &   $9.17^{+1.66}_{-1.23}$ &  $812^{+181}_{-157}$ &             0.98 \\
1383279090527227264$^\mathrm{M}$$^,$$^\mathrm{B}$ &    (65.46, 48.85) &  0.147 $\pm$ 0.016 &   -25.76 $\pm$ 0.03 &   -9.75 $\pm$ 0.04 &  -180.9 $\pm$ 2.4 &   $6.44^{+0.66}_{-0.57}$ &   $8.95^{+0.33}_{-0.25}$ &    $658^{+86}_{-74}$ &             0.98 \\
5932173855446728064$^\mathrm{M}$$^,$$^\mathrm{B}$ &    (329.94, -2.7) &  0.483 $\pm$ 0.029 &    -2.68 $\pm$ 0.04 &   -4.99 $\pm$ 0.03 &  -614.3 $\pm$ 2.5 &   $2.11^{+0.14}_{-0.13}$ &    $6.47^{+0.1}_{-0.11}$ &      $746^{+3}_{-3}$ &             1.00 \\
		\hline
	\end{longtable}
\end{center}
\end{document}